\documentclass[sigconf]{acmart}

\usepackage{algorithmic}
\usepackage{graphicx}
\usepackage{textcomp}
\usepackage{xcolor}
\usepackage{multirow}
\usepackage{threeparttable}
\usepackage{booktabs}     
\usepackage{caption}
\usepackage{tabularx}  
\usepackage{dblfloatfix}
\usepackage{enumitem}
\usepackage{soul}
\usepackage{booktabs}
\usepackage{makecell} 

\newif\ifshowcomments
\showcommentstrue

\AtBeginDocument{%
  }

\copyrightyear{2026}
\acmYear{2026}
\setcopyright{cc}
\setcctype{by}
\acmConference[DAC '26]{63rd ACM/IEEE Design Automation Conference}{July 26--29, 2026}{Long Beach, CA, USA}
\acmBooktitle{63rd ACM/IEEE Design Automation Conference (DAC '26), July 26--29, 2026, Long Beach, CA, USA}
\acmDOI{10.1145/3770743.3804279}
\acmISBN{979-8-4007-2254-7/2026/07}

\begin{document}

\title[Overmind NSA: A Unified Neuro-Symbolic Computing Architecture]{Overmind NSA: A Unified Neuro-Symbolic Computing Architecture with Approximate Nonlinear Activations and Preemptive Memory Bypass}

\author{Weilun Wang}
\affiliation{%
  \institution{University of California, Riverside}
  \city{Riverside}
  \state{CA}
  \country{USA}
}
\email{wwang346@ucr.edu}
\author{Zirui Wang}
\affiliation{%
  \institution{University of California, Riverside}
  \city{Riverside}
  \state{CA}
  \country{USA}
}
\email{zwang829@ucr.edu}
\author{Wantong Li}
\affiliation{%
  \institution{University of California, Riverside}
  \city{Riverside}
  \state{CA}
  \country{USA}
}
\email{wantong.li@ucr.edu}

\begin{abstract}
Neuro-symbolic AI is gaining traction in domains such as large language models, scientific discovery, and autonomous systems due to its ability to combine perception with structured reasoning. However, its deployment is often constrained by high memory demands, diverse computation patterns, and complex hardware requirements. Existing hardware platforms struggle with large on-chip memory overheads, frequent pipeline stalls, limited I/O bandwidth, and inefficient handling of nonlinear operations.
To address these key computational bottlenecks, we propose Overmind, a unified neuro-symbolic architecture with cross-layer optimizations. Overmind tackles these core bottlenecks through Padé approximations for universal nonlinear functions, preemptive memory bypass that eliminates costly on-chip caches, and a complete software stack that optimizes model deployment.
By reconfiguring the Padé orders for approximating nonlinear functions, we also demonstrate adaptive accuracy-performance scaling. Overmind achieves an energy efficiency of 8.1 TOPS/W and a throughput of 410 GOPS for mixed neuro-symbolic workloads with minimal model accuracy loss. Compared to existing solutions, Overmind improves performance and efficiency with significantly fewer hardware resources. 
\end{abstract}




\maketitle

\section{Introduction}
\begin{figure}[b!]
    \centering
    \vspace{-0.2in}
\includegraphics[width=1\linewidth]{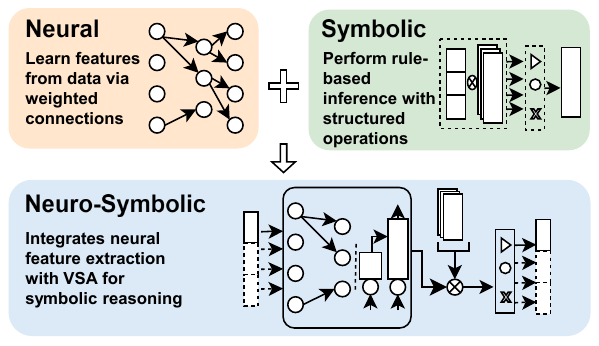}
\caption{Integration of neural and symbolic components, enabling both data-driven learning and rule-based inference.}
     \vspace{-0.05in}
    \label{fig:ns workflow}
\end{figure}

Artificial intelligence (AI) applications increasingly demand models that combine high accuracy with transparency and explainability. To achieve true explainability, researchers aim to embed a reasoning process that is both transparent and traceable, such that any conclusion can be reliably traced back to its foundational data and logical steps. This pursuit has led to the emergence of neuro-symbolic AI (NSA), which unifies two complementary paradigms: connectionism and symbolism~\cite{garcez2023neurosymbolic,NSinAI}.
Connectionist approaches are exemplified by neural networks and excel at pattern recognition and perception, but often require large training datasets and offer limited interpretability. Symbolic approaches encode knowledge through formal symbols and rules for precise and transparent inference, but lack the ability to directly learn from data. By merging these approaches, NSA provides both powerful learning and interpretable decision-making, as illustrated in Fig.~\ref{fig:ns workflow}.

NSA has gained traction across diverse domains, ranging from large language 
models (LLMs) to complex scientific reasoning~\cite{NSAI2024wan,NSprogram,
NSreasoning,NSVision}. 
In safety-critical applications such as robotics and healthcare, this combination is valuable as NSA enables not only accurate predictions but also transparent reasoning that can be inspected and trusted~\cite{lu2024surveying}.
However, these advances introduce additional computational requirements, including memory-bound symbolic operations, intensive matrix similarity searches for codebook lookups, and circular convolution for high-dimensional symbolic binding~\cite{challenges}. 
Conventional GPU and NPU architectures are optimized primarily for linear tensor operations, but will achieve low utilization rate for these memory-intensive workloads. 
While recent NSA accelerators have addressed specific operations, they lack unified support for nonlinear functions, suffer from memory hierarchy inefficiencies, and require specialized hardware for circular convolution that scales poorly with vector dimension~\cite{wan2025cogsys, nsflow}.
In this work, we propose Overmind NSA, a unified neuro-symbolic computing architecture that addresses these limitations through integrated hardware-software co-design. 
\label{sub:NS characterizaion}

\clearpage
Our main contributions are as follows:

\begin{itemize}[leftmargin=*]
\item We design an integrated nonlinear compute engine using Padé approximation that decomposes activation functions into rational polynomials computable directly within PE arrays.
\item We propose preemptive memory bypass with dual-window address filtering that eliminates L2 cache overhead and inherently supports circular convolution via address offset computation.
\item We develop a software stack with co-optimization workflow that automatically maps NSA models to Overmind, selects Padé orders based on accuracy/efficiency targets, and enables deployment of new NSA models without hardware redesign.
\end{itemize}

\section{Background}
\subsection{NSA Algorithms and Operations}
\label{sub:NS operations}

\begin{table}[!t]
\centering
\caption{Fine-Grained Analysis of NSA Atomic Instructions}
\vspace{-3mm}
\label{tab:fine_grained_instructions}
\renewcommand{\arraystretch}{1.2}
\setlength{\tabcolsep}{6pt}
\small
\begin{tabularx}{\linewidth}{@{\extracolsep{\fill}} l|X@{}}
\toprule
\textbf{Model} & \textbf{Compute Type \& Atomic Instruction Composition} \\
\midrule

\textbf{NVSA\cite{NVSA}} &
Vector (98.7\%): elem-wise add, mul, similarity search. 
NN (1.3\%): MAC, ReLU (cmp+mul). \\



\midrule
\textbf{LTN\cite{LTN-badreddine2022logic}} &
Logic (15\%): fuzzy AND, OR, NOT, Element-wise. 
NN (85\%): MAC, smooth activations (sigmoid, tanh). \\

\midrule
\textbf{LNN\cite{LNN-riegel2020logical}} &
Logic (10\%): differentiable AND, OR, IMPLIES, nonlinear (power, sqrt, exp). 
NN (90\%): MAC, sigmoid/ReLU. \\

\midrule
\textbf{NLM\cite{NLM-dong2019neural}} &
NN (55\%): MAC, nonlinear ReLU/tanh/sigmoid. 
Logic (45\%): rule composition, fuzzy AND/OR, MatMul. \\

\bottomrule
\end{tabularx}
\vspace{-5mm}
\end{table}

A typical NSA model consists of two main components: the neural part and the symbolic part. They all share a common architectural pattern, where a neural component first extracts features from tensor data using operations such as convolutions, activations, and fully connected layers. The symbolic component then performs reasoning through operations such as circular convolution, rule detection, and rule execution. 
We further categorize them into vector-symbolic models and logic machines. Vector-symbolic models encode concepts as high-dimensional vectors and perform reasoning through algebraic operations.
For instance, NVSA tackles abstract visual reasoning to infer missing patterns from visual context~\cite{NVSA}.
On the other hand, logic machines represent knowledge through differentiable logical constraints and address knowledge-intensive tasks.
LTN is a popular logic machine that integrates logical constraints into semantic image interpretation~\cite{LTN-badreddine2022logic}. 

\begin{figure}
    \centering
    \includegraphics[width=1\linewidth]{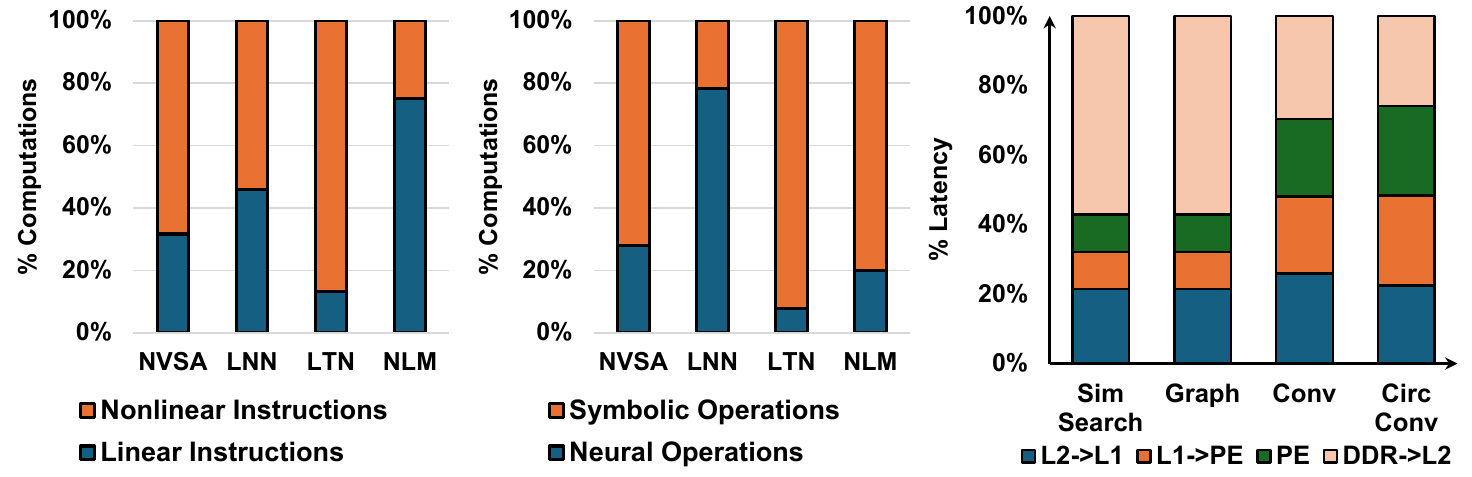}
    \vspace{-0.2in}
    \caption{NSA workload profiling on (a) linear vs. nonlinear computations, (b) neural vs. symbolic operations, and (c) latency breakdown.}
    \vspace{-0.1in}
    \label{fig:percentage}
\end{figure}

NSA models use diverse operations for their neural and symbolic parts, so it is essential to first characterize their computational requirements. We conduct operator-level profiling using PyTorch profiler on multiple NSA models. For each model, we collect cycle-accurate operation counts and data 
access patterns. Table~\ref{tab:fine_grained_instructions} summarizes the fundamental operations for four representative models, which cover common NSA operations and represent both vector-symbolic and logic-based paradigms.
The diversity of neuro-symbolic operations indicates that existing general-purpose architectures cannot efficiently support the broad spectrum of NSA workloads.
Our analysis in Fig.~\ref{fig:percentage}a reveals that nonlinear operations dominate the computations in multiple NSA models. The symbolic components also require comparable or even more resources than the neural components, as shown in Fig.~\ref{fig:percentage}b. These trends confirm that traditional accelerator strategies, such as scaling MAC units and optimizing for neural operators, are insufficient for NSA workloads. 
As shown in Fig.~\ref{fig:percentage}c, symbolic operations exhibit more memory-bound access patterns and lead to higher memory transfer latencies compared to neural operations.

\vspace{-0.1in}
\subsection{Existing NSA Computation Challenges}
\label{sub:NS hardware}

\begin{figure}[t!]
    \centering
    \includegraphics[width=1\linewidth]{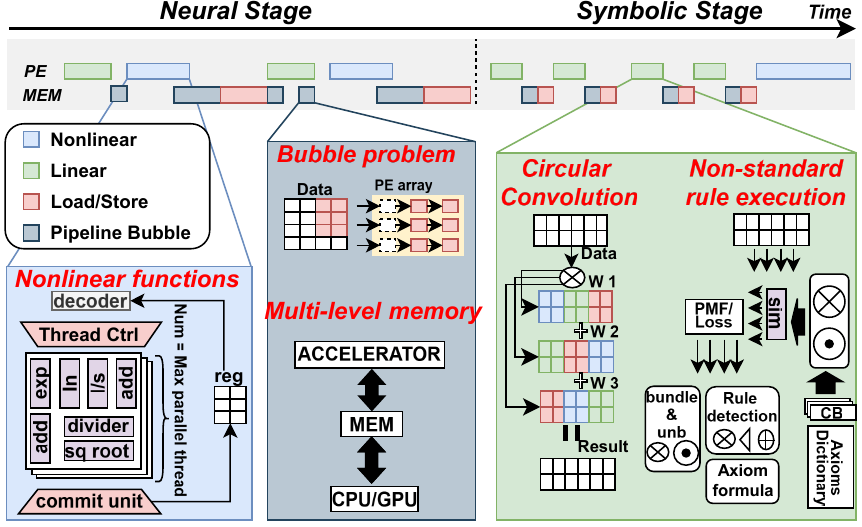}
    \caption{Summary of key computational bottlenecks in neural and symbolic components of NSA models.}
    \label{fig:bottleneck}
    \vspace{-0.2in}
\end{figure}

While existing hardware accelerators have been effective for parallelized linear operations in neural networks, they cannot fully address the unique properties of NSA workloads. Existing designs based on processing engine (PE) arrays~\cite{TPU,eyeriss,mtia,pearray, dave2021hardware} and memory-centric architectures~\cite{ahn2015scalable, li2023h3datten, mutlu2024memory, gebregiorgis2022survey} are optimized for neural operations, but achieve low utilization for symbolic computation patterns such as circular convolution and rule-based reasoning. Fig.~\ref{fig:bottleneck} summarizes the key computational bottlenecks when executing NSA models on general-purpose hardware.

\noindent\textbf{Architectural mismatch for nonlinear computations.}
NSA workloads devote a large fraction of execution cycles to nonlinear functions, yet existing accelerators provide limited hardware support for these primitives. Existing approximation methods introduce architectural bottlenecks. Taylor series approximations suffer from limited convergence radius and require piecewise implementations with multiple expansion points that increase control overhead and storage requirements~\cite{sharma2017activation}. Lookup tables (LUTs) consume significant on-chip memory and impose a heterogeneous dataflow, where linear operations use PE arrays while nonlinear operations require separate access, leading to underutilized PE resources~\cite{wuraola2022resource}. Iterative methods introduce variable latency that disrupts pipelining~\cite{armeniakos2022hardware}. These methods all share a fundamental limitation of treating nonlinear computation as a post-processing step. To maintain both high hardware utilization and low latency, nonlinear computations must be directly enabled within the PE array.

\noindent\textbf{Memory hierarchy inefficiency for symbolic operations.} Symbolic operations in NSA exhibit memory-bound access patterns that underutilize conventional memory hierarchies. For example, codebook searches exhibit large-stride access with working sets that can exceed cache capacity. These patterns achieve low cache hit rates compared to neural convolutions with favorable locality~\cite{NSAI2024wan}. Multi-level hierarchies exacerbate this inefficiency: L2→L1 data transfers during inter-layer transitions cause pipeline stalls, an effect worsened in NSA workloads that alternate between neural layers and symbolic operations lacking temporal reuse. For symbolic operations with poor hit rates, this hierarchy adds latency without performance benefits. An effective solution must replace cache-based speculation to eliminate L2 overhead while maintaining high memory bandwidth.

\noindent\textbf{Inefficient circular convolution for symbolic binding.}
Circular convolution is a core primitive in NSA architectures for binding high-dimensional vectors~\cite{NSAI2024wan}. Prior hardware accelerators implement this operation using staggered shift registers across PE lanes~\cite{kleyko2022vector}. However, their area costs scale poorly with vector dimension, and they serialize data delivery so PEs must wait for rotated data to propagate through shift chains and thus suffer from pipeline bubbles. This occurs due to decoupling circular convolution from the main PE datapath, which reduces overall utilization. A scalable solution must eliminate dedicated rotation hardware by supporting circular convolution through dynamic index remapping derived from operator metadata such as stride and shift amount.

\begin{figure}
    \centering
    \includegraphics[width=1\linewidth]{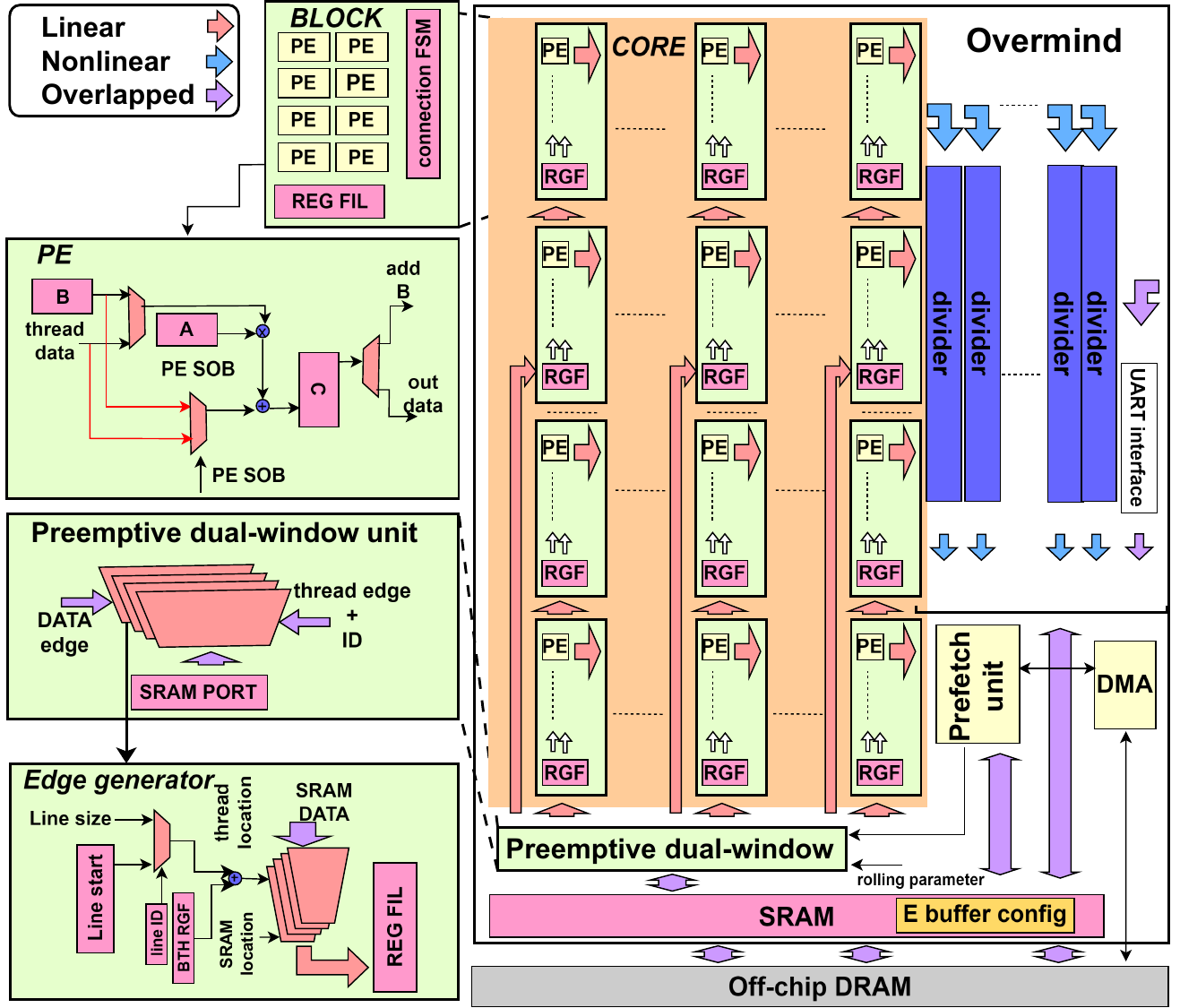}
    \caption{The Overmind NSA hardware architecture.}
    \label{fig:architecture}
    \vspace{-0.2in}
\end{figure}

\section{Overmind NSA Architecture}

To address the computational challenges of NSA workloads, we introduce the Overmind architecture for unified high-performance execution of hybrid neuro-symbolic workloads. As illustrated in Fig.~\ref{fig:architecture}, Overmind integrates four key subsystems: a reconfigurable PE array, a high-throughput divider array, a dual-window preemptive unit for memory address filtering, and a streamlined single-level memory subsystem. Together, they provide three core capabilities: (i) scalable approximate execution for complex nonlinear functions, (ii) high-throughput symbolic processing with memory-efficient dataflows, and (iii) custom model deployment with compile-time accuracy-performance optimization via adaptive Padé configuration. Overmind accounts for the unique characteristics of NSA workloads and is tailored for models that combine neural computation with symbolic reasoning. It supports both standard neural primitives and NSA-specific symbolic operations. The compiler-automated model deployment allow new NSA workloads to be instantiated on Overmind without hardware redesign.

\vspace{-0.08in}
\subsection{Approximate Nonlinear Compute Engine}

\begin{figure*}
    \centering
    \includegraphics[width=0.95\textwidth]{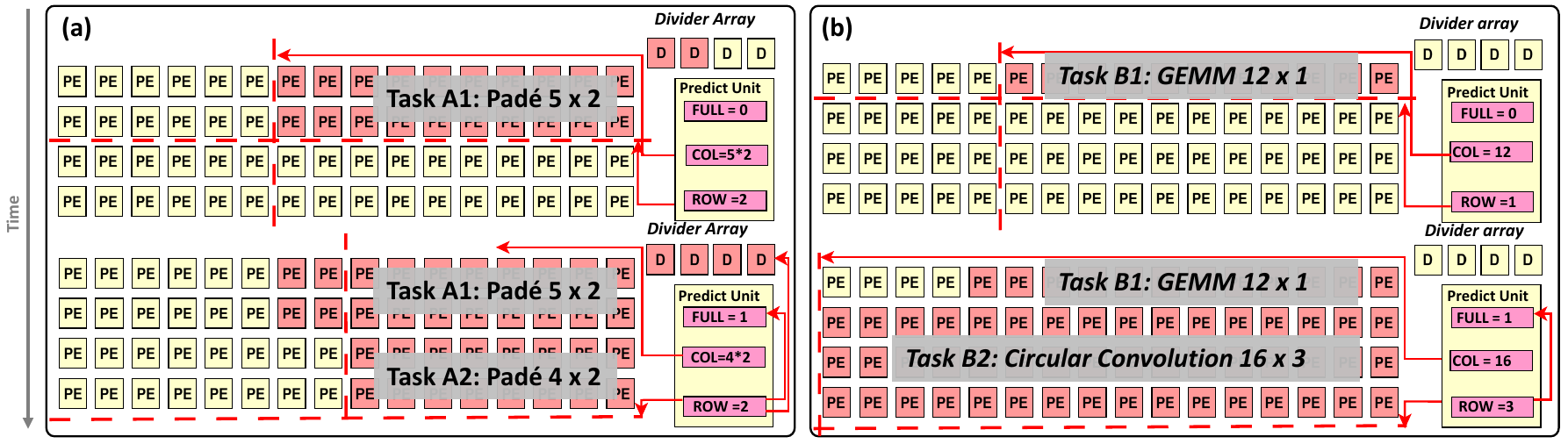}
    \caption{Reconfigurability of Overmind PE array across neural and symbolic 
    operations. Divider arrays are enabled for approximate computations (a), but otherwise deactivated (b). Red and yellow denote active and idle units respectively.}
    \label{fig:PE}
    \vspace{-0.1in}
\end{figure*}

Nonlinear functions account for significant computations in NSA workloads, yet existing techniques for executing nonlinear functions
exhibit poor architectural alignment with conventional hardware~\cite{sharma2017activation, armeniakos2022hardware}. These methods require dedicated post-processing logic and cannot be efficiently mapped to regular PE arrays, leading to underutilization and increased latency.
We address this challenge using Padé approximation, which expresses nonlinear functions as a ratio between two polynomials (Eq.~\ref{eq:Padé}).
Padé approximants capture global function behaviors with fewer terms than Taylor series and allow more accurate approximations over wider input ranges~\cite{brezinski1994pade}. Crucially, both the Padé numerator and denominator consist of multiply-accumulate (MAC) operations, which can be computed directly in Overmind PE array with deterministic latency.

\begin{equation}
    \label{eq:Padé}
R_{m,n}(x) = \frac{\sum_{i=0}^{m} a_i x^i}{1 + \sum_{j=1}^{n} b_j x^j}
\end{equation}

Overmind implements Padé approximation by augmenting its PE array with integrated divider units. The PE array evaluates the numerator and denominator polynomials in parallel, while the dividers complete the final rational function. This avoids the memory and latency overheads of LUT-based methods, which require high-resolution tables and interpolation hardware. In Overmind, the PE array is organized as $R$ rows $\times$ $C$ columns, where each row processes one thread. To avoid division bottlenecks, we allocate one divider per row to allow $R$ parallel divisions. 
The PE datapath incorporates two mechanisms for efficient Padé execution. 
First, exponent accumulation reuses internal registers to compute powers of $x$ ($x^i$) through chained multiplications and is fused with coefficient multiplication in the same MAC pipeline. This eliminates dedicated exponentiation hardware while maintaining single-pass evaluation. 
Second, coefficient broadcasting preloads Padé coefficients $(a_i, b_i)$ from shared SRAM and distributes them to active PEs for parallel execution across multiple data elements without per-thread fetch overhead. Since coefficients are constant for a given approximation, this broadcast mechanism maximizes SRAM bandwidth utilization.

Fig.~\ref{fig:PE} illustrates Overmind's ability to reconfigure its PE array for diverse NSA operations. 
For Padé approximation, Overmind activates PE columns proportional to the polynomial order. 
For instance, Padé-5 approximants use 10 (5 for numerator, 5 for denominator) columns with 2 threads. 
For linear operations, the dividers remain idle and the PE array operates in standard mode.
Multiple data elements can be processed in parallel by assigning each to a different row, with each row operating as an independent thread. 
This unified reconfigurable architecture enables both neural and symbolic NSA stages to efficiently execute in parallel.
In addition, the reconfigurability supports compile-time accuracy-performance optimization through Padé order selection. Higher-order approximants activate more PE columns and offer better precision, 
whereas lower orders improve system throughput at the cost of minor accuracy degradations. As detailed in Section 3.3, users can specify target accuracy through compiler directives, and our framework automatically selects the appropriate Padé orders and synthesizes the corresponding hardware mapping.

\vspace{-0.1in}
\subsection{Preemptive Memory Bypass}
\label{sec:memory}

\begin{figure*}
\centering
\includegraphics[width=1\textwidth]{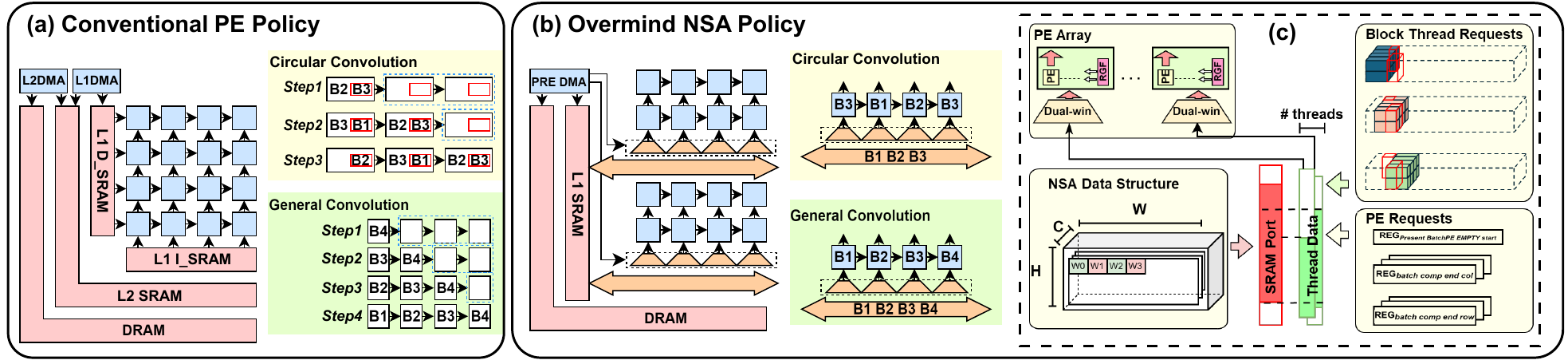}
\caption{(a) Conventional PE policy and memory hierarchy with L2$\rightarrow$L1 transfers. (b) Overmind's preemptive memory bypass and broadcast-based bypass design. (c) Dual-window filter logic for 2D address range filtering.}
\label{fig:expenation}
\vspace{-0.1in}
\end{figure*}

Symbolic operations in NSA exhibit memory-bound access patterns such as large strides in codebook search and modulo indexing in circular convolution, which conventional cache hierarchies fail to exploit. Two-level SRAM systems rely on explicit L2$\rightarrow$L1 transfers that introduce pipeline stalls, especially during frequent transitions between neural and symbolic layers (Fig.~\ref{fig:expenation}a). 
For NSA workloads dominated by symbolic operations, L2 caching provides little benefit and adds latency.
We propose a preemptive bypass mechanism effective for symbolic operations by deriving access patterns from tensor metadata.
Overmind bypasses the L2 cache and instead computes valid 2D address ranges directly from tensor metadata. Each instruction carries parameters such as shape, stride, and kernel size. The controller uses these to configure address windows for each PE row before execution. During execution, SRAM broadcasts data to all rows over a shared bus. Each row uses a dual-window comparator, one window for the row dimension and one for the column dimension, to select relevant elements from the broadcast stream based on its configured window (Fig.~\ref{fig:expenation}b).
To minimize broadcast power, the controller generates a row enable mask from address window metadata and activates only the PE rows whose windows intersect 
the current broadcast address range.

Fig.~\ref{fig:expenation}c shows the local filter logic in each PE row. It maintains three registers (batch start, row boundary, and column boundary) that together define the valid address range for the current output element. Incoming address tags are compared against this range to select matching data for computation. This mechanism adds minimal overhead and removes the need for large L2 SRAM buffers.
This bypass mechanism also supports circular convolution without shift registers. Traditional designs rotate operands across PE columns, which does not scale for long vectors. Overmind avoids this by applying a circular offset to each row's address window. For $C[i] = \sum_{j=0}^{N-1} A[j] \times B[(i-j) \bmod N]$, PE $i$ sets its address window to start at $(\text{base} + i) \bmod N$. The dual-window logic selects wrapped-around values directly from the stream to avoid dedicated shift register hardware. 
Finally, the controller pre-decodes the next instruction and loads window values into each row before the current layer completes, which hides latency and maintains throughput. By eliminating L2 cache and overlapping memory reads, Overmind also resolves the pipeline bubble issues in conventional architectures.

\begin{figure}
    \centering
    
    \includegraphics[width=1\linewidth]{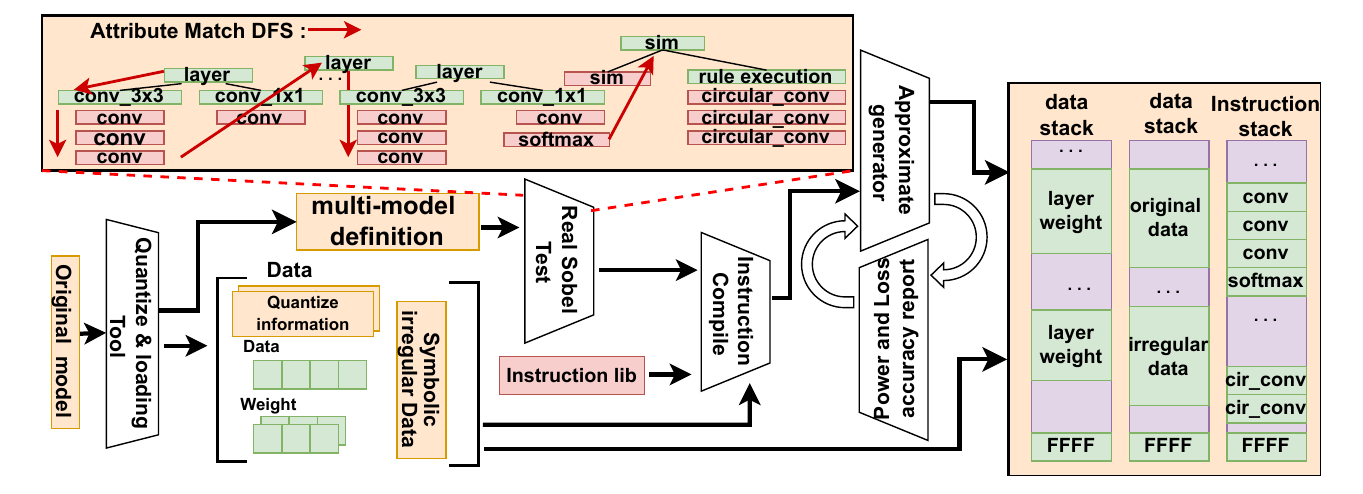}
    \caption{Software stack with hardware co-optimization.}
    \label{fig:software workflow}
    \vspace{-0.1in}
\end{figure}

\vspace{-0.1in}
\subsection{Overmind Software Workflow}
We develop a complete software stack for deploying diverse NSA models on Overmind. Unlike traditional compilation to fixed hardware, our approach exploits Overmind's reconfigurability through hardware-software co-optimization. The compiler analyzes model requirements and generates both instructions and hardware configuration parameters, while the runtime system dynamically reconfigures PE arrays, memory policies, and power states based on per-layer characteristics, as depicted in Fig.~\ref{fig:software workflow}.

\noindent\textbf{Model compilation with co-optimization.} To support multi-model execution~\cite{multi-model}, the compiler performs an attribute-based depth-first search (DFS) that traverses the model's computational graph and extracts operator types, tensor shapes, and data dependencies. The DFS identifies both standard neural operations and NSA-specific symbolic operations. This creates a unified intermediate representation that abstracts away framework-specific implementations. During this analysis, the compiler extracts computational characteristics (nonlinearity distribution, memory access patterns, symbolic operation density) that guide downstream hardware configuration decisions.

\noindent\textbf{Adaptive instruction generation.} The instruction generator performs hardware-software co-optimization by mapping each operator to both an instruction and a hardware configuration. For nonlinear activation functions, the adaptive approximation generator synthesizes Padé approximants with order $(m,n)$ jointly optimized between software requirements and hardware constraints. Users specify accuracy targets via compiler flags, and the compiler iteratively optimizes for the minimum Padé order that satisfies the constraint. For each layer, the compiler generates an instruction bundle containing the operation encoding, hardware configuration parameters, and additional metadata. This co-designed instruction format allows the runtime system to reconfigure hardware per-layer without additional analysis overhead.

\noindent\textbf{Runtime execution with hardware reconfiguration.} The runtime system organizes the compiled model into execution stacks for data, instructions, and intermediate results. During execution, the activation controller performs dynamic hardware reconfiguration based on instruction metadata. The controller implements a memory placement policy determined at compile-time, where tensor metadata embedded in instructions guides optimal data placement between SRAM and DDR. Power co-optimization occurs through selective activation of PE columns, dividers, and memory banks based on per-operation requirements.

\section{Evaluation}

\subsection{Software-Level Evaluation}

\begin{figure}
    \centering
\includegraphics[width=1\linewidth]{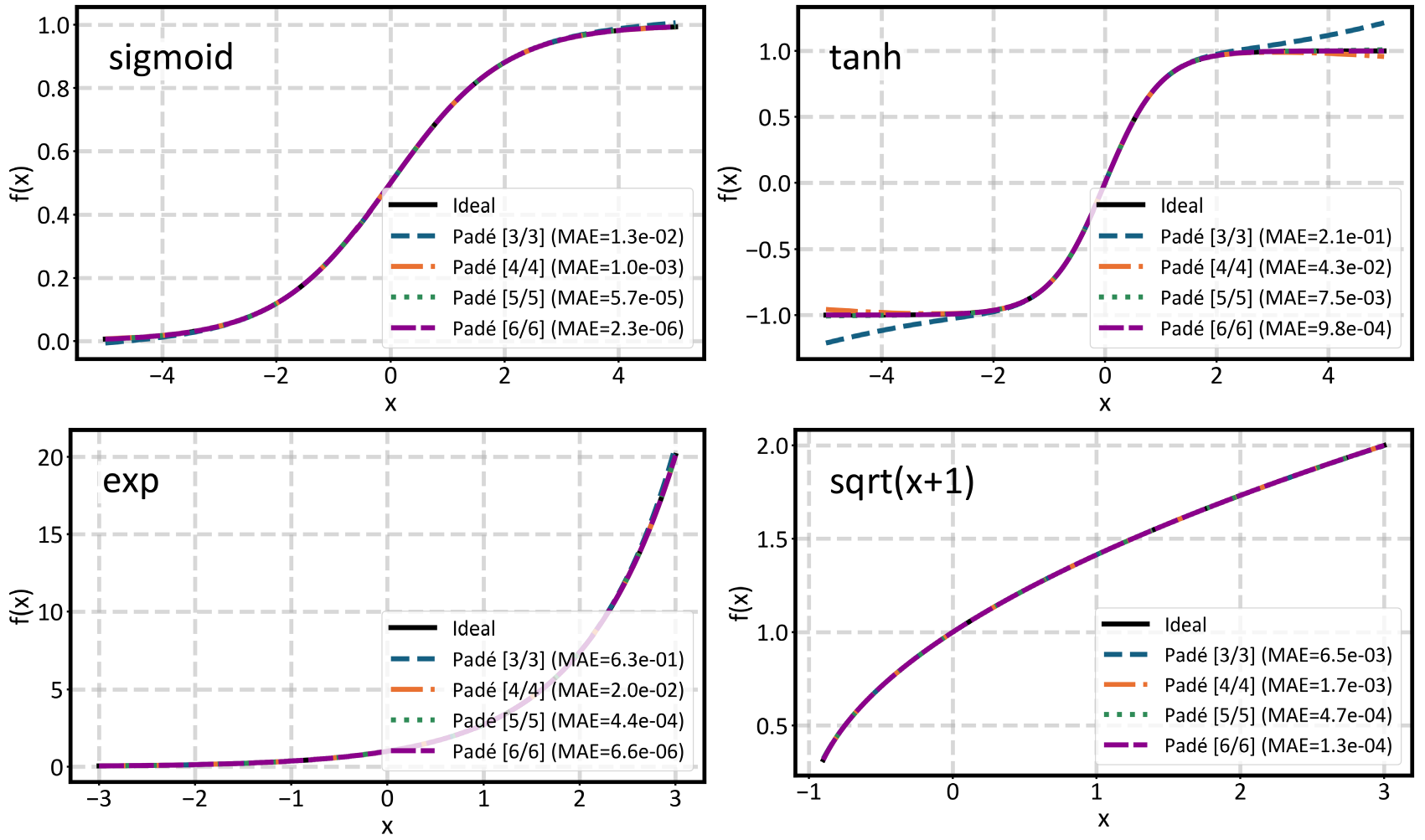}
    \caption{Comparison between ideal nonlinear functions and Padé-approximated versions.}
    \label{fig:approximate accuracy}
    \vspace{-0.1in}
\end{figure}

\begin{table}
\centering
\caption{Model Accuracy Loss under Different Padé Orders}
\vspace{-2mm}
\label{tab:accuracy}
\small
\begin{tabular}{|c||c|c|c|c|}
\hline
\textbf{Accuracy Loss} & \textbf{Padé-3} & \textbf{Padé-4} & \textbf{Padé-5} & \textbf{Padé-6} \\
\hline

\begin{tabular}{@{}c@{}}Baseline NVSA (\%)\end{tabular} & 0.61 & 0.11 & 0 & 0 \\
\cline{1-5}
\begin{tabular}{@{}c@{}}INT8 NVSA (\%)\end{tabular} & 0.7 & 0 & 1.02 & 0 \\
\cline{1-5}
\begin{tabular}{@{}c@{}}Baseline NLM (\%)\end{tabular} & 0 &  0 & 0 & 0 \\
\cline{1-5}
\begin{tabular}{@{}c@{}}INT8 NLM (\%)\end{tabular} & 2.96 & 0.74 & 0.74 & 0 \\
\cline{1-5}
\begin{tabular}{@{}c@{}}Baseline LTN (\%)\end{tabular} & 0 & 0 & 0 & 0 \\
\cline{1-5}
\begin{tabular}{@{}c@{}}INT8 LTN (\%)\end{tabular} & 1.96 & 0 & 0.392 & 0.392 \\
\hline

\end{tabular}
\vspace{-0.2in}
\end{table}

\noindent\textbf{Models and datasets.} 
To evaluate cognitive reasoning capability, we conduct experiments on the commonly used spatial-temporal reasoning RAVEN and I-RAVEN datasets, which measure abstract visual reasoning through progressive matrix completion tasks~\cite{raven, IRAVEN}. These datasets are widely adopted for NSA evaluation as they require both perceptual feature extraction (neural stage) and rule-based reasoning (symbolic stage).
We select NVSA, NLM, and LTN models as representative NSA workloads to evaluate Overmind, as these models comprise of all common neural and symbolic operators~\cite{NVSA, NLM-dong2019neural, LTN-badreddine2022logic}. 
The NSA models are trained on an RTX 2080 Ti GPU. For all models, we apply post-training quantization to INT8 format for both activations and model weights. 

\noindent\textbf{Approximate nonlinear functions.}
For the proposed Overmind algorithms, we analyze the accuracy of approximate nonlinear functions.
We evaluate the precision of our approximation method using the maximum absolute error (MAE) metric across multiple common non-linear functions. 
To ensure broad applicability, we also test custom functions such as square root, which are used in both NSA and transformer models~\cite{vaswani2017attention}. 
As summarized in Fig.~\ref{fig:approximate accuracy}, our method maintains low errors across a wide input range. We also observe higher Padé order leads to improvement in MAE.

\noindent\textbf{End-to-end model accuracy.} 
Next, we evaluate the resulting NSA reasoning accuracy when all nonlinear functions in each NSA model use the Padé approximation.
With the considered RAVEN and I-RAVEN datasets, the average accuracy loss also reduces with higher Padé orders (Table~\ref{tab:accuracy}), which forms the basis for our adaptive accuracy-performance scaling.

\subsection{Hardware Evaluation}

\begin{figure}
    \includegraphics[width=1\linewidth]{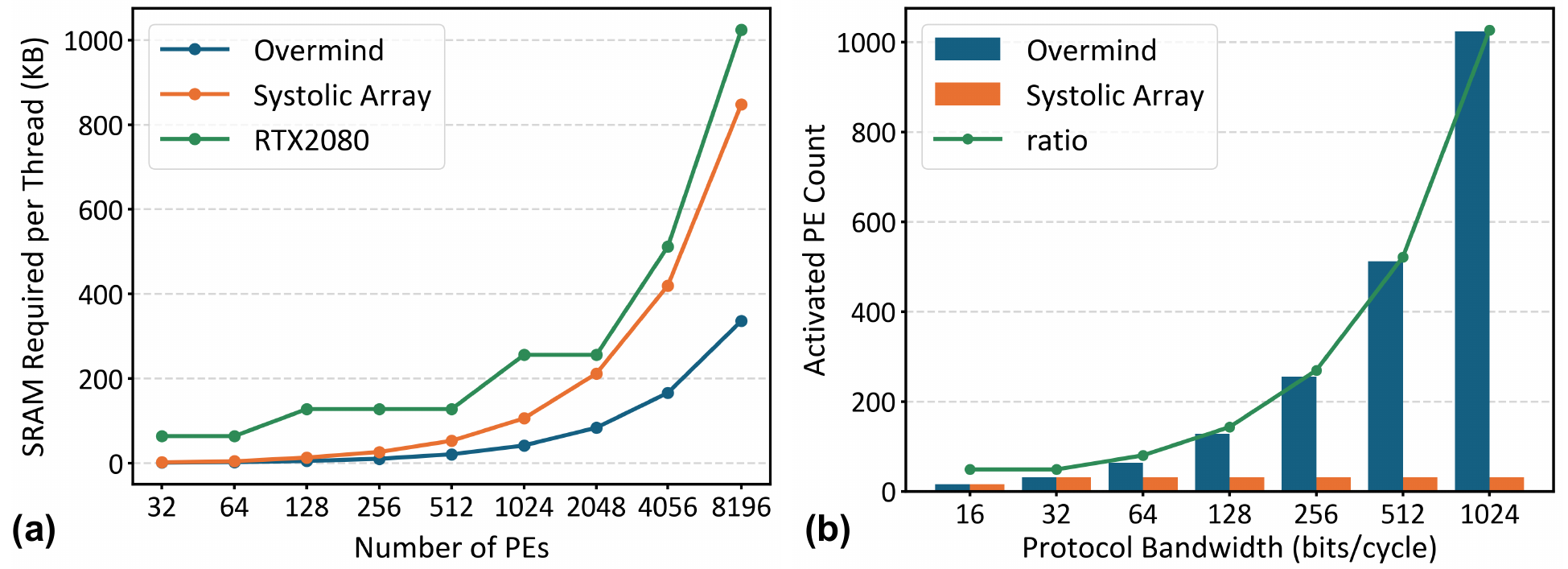}
    \caption{Scalability analyses: (a) Overmind requires fewer SRAM as number of PEs increases and (b) Overmind can activate more PEs given the same bandwidth.}
    \label{fig:memoryincrease}
    \vspace{-3mm}
\end{figure}

\noindent\textbf{ASIC design.}
We first evaluate Overmind using a standard ASIC design flow.
The design is implemented in RTL and synthesized with Synopsys Design Compiler using GlobalFoundries 22-nm process. Power estimates are obtained using Synopsys PrimeTime.
The configuration used for synthesis consists of 32 rows by 16 columns of PEs and 32 KB of on‑chip SRAM. The design delivers an average throughput of 410 GOPS at 800 MHz for mixed neuro-symbolic workloads.
Table~\ref{tab:ppa_breakdown} reports the post-synthesis area and power breakdown with Padé order configured to 4.
The approximate function units contribute 2.6\% power and 5.7\% area overheads.
To assess accuracy–performance trade-offs, we evaluate the NVSA model across multiple Padé configurations. Reducing the Padé order from 6 to 3 improves throughput by approximately 2$\times$, but with a 0.6\% loss in model accuracy. With this trade-off, Overmind’s software stack adjusts the approximation level to meet user-defined accuracy requirements.
The preemptive memory bypass introduces additional hardware for window filtering, which accounts for 14.5\% power and 8\% area. These costs are offset by the removal of the on‑chip L2 cache, saving around 0.84 mm$^2$ of SRAM area for Overmind. The bypass mechanism also improves the hardware scalability. As shown in Fig.~\ref{fig:memoryincrease}, Overmind requires less SRAM per thread than systolic or GPU‑style architectures as the number of PEs increases, and it sustains higher PE utilization at the same memory bandwidth.
With Padé=4, Overmind achieves an energy efficiency of 8.1 TOPS/W, which is 40.5$\times$ higher than a Xeon CPU and 8$\times$ higher than an RTX GPU.
Table~\ref{tab:hw_comparison} compares Overmind with existing PE-based accelerators, including TPU \cite{TPU}, MTIA \cite{mtia}, and CogSys\cite{wan2025cogsys}. For a fair comparison, all designs are normalized to the same process technology and PE count. Overmind demonstrates the highest speed and highest energy efficiency for full NSA workloads.

\begin{table}[!t]
\centering
\caption{Area and power breakdowns of Overmind}
\vspace{-3mm}
\label{tab:ppa_breakdown}
\small
\begin{tabular}{|c||c|c|}
\hline
\textbf{Component} & \textbf{Area (mm\textsuperscript{2})} & \textbf{Power (mW)} \\
\hline
PE Array        & 0.1005 & 37.52 \\
Divider         & 0.0083 & 1.33 \\
Memory Bypass   & 0.0117 & 7.32 \\
Cache           & 0.0262 & 4.46\\
\hline
\textbf{Total}  & \textbf{0.1467} & \textbf{50.63} \\
\hline
\end{tabular}
\end{table}




\begin{table}
\centering
\caption{Comparison with PE-based platforms}
\vspace{-3mm}
\label{tab:hw_comparison}
\small
\begin{tabular}{|c||c|c|c|c|}
\hline
\textbf{Property} & \textbf{TPU} & \textbf{MTIA} & \textbf{CogSys} & \textbf{Overmind} \\
\hline


PE Config. & 1 128$\times$128 & 16 32$\times$32 & 64 16$\times$16 & 16 32$\times$32 \\
\cline{1-5}


\begin{tabular}{@{}c@{}}Norm. Speed\end{tabular} & 1.0$\times$ & 1.91$\times$ & 7.39$\times$ & 12.26$\times$ \\
\cline{1-5}
\begin{tabular}{@{}c@{}}Norm. Energy\\Efficiency \end{tabular} & 1.0$\times$ & 1.6$\times$ & 27.02$\times$ & 27.55$\times$ \\
\hline
\end{tabular}
\end{table}

\begin{table}
\centering
\caption{Comparison with FPGA-based NSA accelerator}
\vspace{-3mm}
\label{tab:design_config_vertical_rotated_fixed}
\small
\begin{tabular}{|c||c|c|}
\hline
\textbf{Property} & \textbf{\begin{tabular}{@{}c@{}}NSFlow\end{tabular}} & \textbf{Overmind} \\
\hline

Data Type & INT8 \& INT4 & INT8 \\
\cline{1-3}
\begin{tabular}{@{}c@{}}PE Config.\end{tabular} & 32$\times$16 & 32$\times$16 \\
\cline{1-3}
\begin{tabular}{@{}c@{}} Norm. Throughput\end{tabular} & 1.0$\times$ & 1.5$\times$ \\

\cline{1-3}
DSPs & 341 & 133 \\
\cline{1-3}
LUTs & 30,240 & 21,419 \\
\cline{1-3}
DFFs & 64,800 & 30,627 \\
\cline{1-3}
Memory & 1.2 MB & 21 KB \\
\hline
\end{tabular}
\vspace{-3mm}
\end{table}

\noindent\textbf{FPGA validation.} 
We also implement Overmind on FPGA to compare with NSFlow, the only existing FPGA-based NSA accelerator~\cite{nsflow}.
The evaluation is performed on a Kintex‑7 FPGA running the NVSA model end‑to‑end~\cite{NVSA}. We use our custom software stack to generate the config and compute data, and program the FPGA via UART. For a fair comparison, we match Overmind’s PE configuration to that of NSFlow. 
As summarized in Table~\ref{tab:design_config_vertical_rotated_fixed}, Overmind reduces on‑device memory by 57.1$\times$ due to the removal of the L2 cache. 
It also uses fewer compute resources because the divider-based approximation is less complex than DSP-based designs. At the same frequency, Overmind achieves 1.5$\times$ higher throughput compared to NSFlow, enabled by its preemptive read mechanism and broadcast‑based data delivery.

\noindent\textbf{Related works.} 
CogSys extends systolic array architectures with specialized support for circular convolution, but lacks unified mechanisms for nonlinear functions and retains multi-level memory hierarchies that underutilize for symbolic operations~\cite{wan2025cogsys}. NSFlow provides an FPGA-based framework with automated dataflow generation for NSA workloads, but does not address the core computational bottlenecks of nonlinear approximation or memory hierarchy inefficiencies~\cite{nsflow}. TinyNS performs neural architecture search to adapt NSA models for existing microcontroller hardware~\cite{saha2024tinyns}. General-purpose DNN accelerators such as TPU and MTIA optimize for dense matrix operations, but are not optimized for symbolic operations and frequent nonlinear activations~\cite{TPU, mtia, pearraysurvey}.

\section{Conclusion}

This work presents Overmind NSA, a unified architecture optimized for end-to-end neuro-symbolic AI workloads. Overmind addresses key computational and memory bottlenecks through three co-optimized innovations: Padé-based nonlinear approximations for reasoning efficiency, a preemptive memory bypass mechanism to eliminate on-chip cache dependencies, and a stall-free parallel dataflow design to maximize PE utilization.
ASIC and FPGA evaluations demonstrate that Overmind delivers substantial improvements in throughput and energy efficiency while preserving symbolic reasoning accuracy. 
With its general-purpose design and adaptive software stack, Overmind offers a programmable and efficient execution platform for future neuro-symbolic applications at scale.

\begin{acks}
This work is supported in part by the Regents' Faculty Fellowship, granted by the UC Riverside Academic Senate.
\end{acks}

\clearpage

\bibliographystyle{ACM-Reference-Format}
\bibliography{sample-base}

\end{document}